\documentclass[a4paper]{article}
\usepackage{amsmath,graphicx}

\usepackage{cite,xcolor}
\usepackage{booktabs} 
\usepackage{multirow}
\usepackage{multicol}
\usepackage{amsfonts}

\usepackage{amssymb}% http://ctan.org/pkg/amssymb
\usepackage{pifont}% http://ctan.org/pkg/pifont
\newcommand{\cmark}{\ding{51}}%
\newcommand{\xmark}{\ding{55}}%
\usepackage{url}
\usepackage[font=small]{caption}

\usepackage{INTERSPEECH2022}

\title{A Hybrid Continuity Loss to Reduce Over-Suppression for \\ Time-domain Target Speaker Extraction}
\name{Zexu Pan$^{1}$, Meng Ge$^{2}$, Haizhou Li$^{1,3,4}$ 
\thanks{This research is supported by  
    The Science and Engineering Research Council, Agency for Science, Technology and Research (A*STAR), Singapore, through the National Robotics Program under Human-Robot Interaction Phase 1 (Grant No. 192 25 00054), 
    by the Deutsche Forschungsgemeinschaft (DFG, German Research Foundation) under Germany's Excellence Strategy (University Allowance, EXC 2077, University of Bremen), and by the Guangdong Provincial Key Laboratory of Big Data Computing, The Chinese University of Hong Kong, Shenzhen under Grant No. B10120210117-KP02, UDF01002333, and UF02002333.}
}
\address{
  $^{1}$Institute of Data Science, National University of Singapore, Singapore\\
  $^{2}$Key Laboratory of Cognitive Computing and Application, Tianjin University, China \\
  $^{3}$The Chinese University of Hong Kong, Shenzhen, China\\
  $^{4}$Machine Listening Lab, University of Bremen, Germany}
\email{pan\_zexu@u.nus.edu, gemeng@tju.edu.cn, haizhou.li@nus.edu.sg}

\begin{document}

\maketitle

\begin{abstract}
  The speaker extraction algorithm extracts the target speech from a mixture speech containing interference speech and background noise. The extraction process sometimes over-suppresses the extracted target speech, which not only creates artifacts during listening but also harms the performance of downstream automatic speech recognition algorithms. We propose a hybrid continuity loss function for time-domain speaker extraction algorithms to settle the over-suppression problem. On top of the waveform-level loss used for superior signal quality, i.e., SI-SDR, we introduce a multi-resolution delta spectrum loss in the frequency-domain, to ensure the continuity of an extracted speech signal, thus alleviating the over-suppression. We examine the hybrid continuity loss function using a time-domain audio-visual speaker extraction algorithm on the YouTube LRS2-BBC dataset. Experimental results show that the proposed loss function reduces the over-suppression and improves the word error rate of speech recognition on both clean and noisy two-speakers mixtures, without harming the reconstructed speech quality.
\end{abstract}

\begin{figure*}
  \centering
  \includegraphics[width=0.98\linewidth]{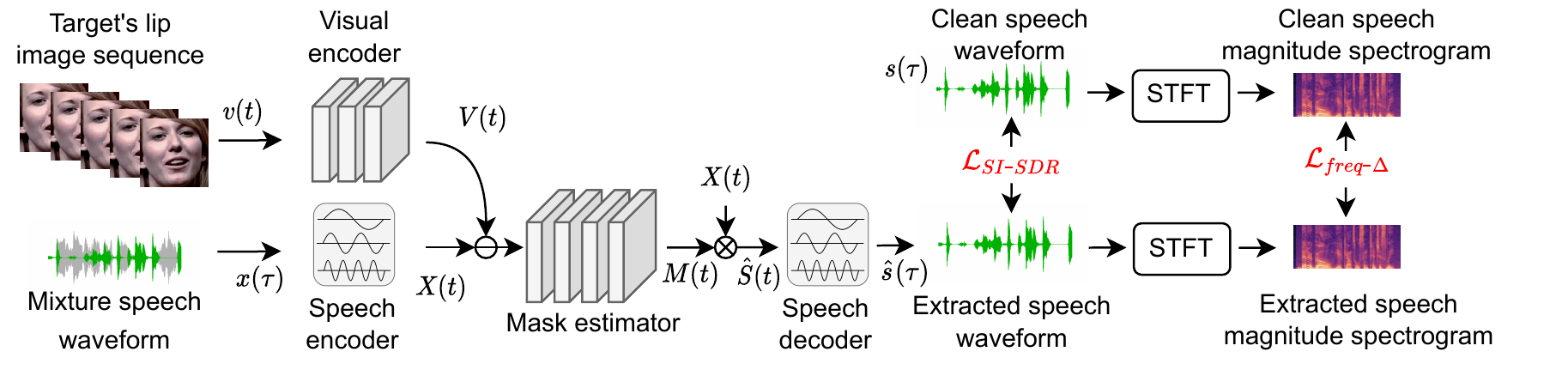}
%   \vspace*{-2mm}
  \caption{The pipeline of a mask-based time-domain audio-visual speaker extraction network. $\ominus$ represents the concatenation of features along the channel dimension and $\otimes$ represents element-wise multiplication. We proposed a hybrid continuity loss function that consists of the time-domain loss $\mathcal{L}_{SI\mbox{-}SDR}$ and frequency-domain delta spectrum loss $\mathcal{L}_{freq\mbox{-}\Delta}$ at different resolutions, to alleviate the over-suppression problem.}
  \label{fig:network}
%   \vspace*{-2mm}
\end{figure*}

\section{Introduction}
The acoustic environment during real-world human-robot interaction can be described as a cocktail party~\cite{bronkhorst2000cocktail}, where the speech from a speaker of interest, i.e., the target speaker, is often corrupted by interference speakers and background noise. In such a scenario, speech separation or speaker extraction algorithms are usually needed to extract the clean speech signal of the target speaker~\cite{ephrat2018looking,reentry}, which is a crucial step for downstream applications such as hearing aid development~\cite{wang2017deep}, active speaker detection~\cite{tao2021someone}, speech recognition~\cite{yue2019end,song2022multimodal}, and source localization~\cite{qian2021multi}.

The speech separation algorithm~\cite{luo2020dual,luo2019conv,kolbaek2017multitalker,hershey2016deep,zeghidour2020wavesplit,liu2019divide} separates a mixture speech signal into individual streams, each containing a clean speech signal of a speaker. Inside the network, multiple speaker streams compete and segregate, which usually requires the number of speakers to be known.

The speaker extraction algorithm takes a different approach, which only extracts the speech of the target speaker from the mixture. It does not need to know the number of speakers but requires an auxiliary reference to distinguish the target speaker from the others. The auxiliary reference can be a pre-recorded reference speech signal, in which the algorithm extracts a speech signal that has a similar voice signature to the reference speech signal~\cite{wang2020voicefilter,Chenglin2020spex,spex_plus2020,xiao2019single,shi2020speaker,delcroix2020improving}. The video recording of the target speaker also serves as such a reference, in which the algorithm extracts a speech signal that is temporally synchronized with the speaker's motion in the video~\cite{wu2019time,afouras2018conversation,ochiai2019multimodal,seg_pan,koichiro2021,michelsanti2021overview,pan2020muse,pan2021usev}.

Time-domain speaker extraction algorithm is preferable to the frequency-domain counterparts because it gets rid of the phase reconstruction problem in Short Time Fourier Transform (STFT)~\cite{luo2019conv,spex_plus2020}. However, it experiences the over-suppress problem in extracting the target signal, in which the time-frequency bins of the extracted speech magnitude spectrogram are attenuated to a lower value than expected~\cite{wang2020voicefilter}. This not only creates auditory artifacts, but also adversely affects downstream applications such as automatic speech recognition (ASR).

Addressing the over-suppression problem, VoiceFilter-Lite~\cite{wang2020voicefilter} took the frequency-domain approach with an asymmetric loss function, which penalizes the over-suppression errors more than the under-suppression errors on the magnitude spectrogram. The asymmetric loss function is designed to improve the ASR performance, but it remains a question whether this will adversely impact the speech reconstruction. In this paper, we seek to alleviate the over-suppression problem in time-domain speaker extraction algorithms, so as to improve the downstream ASR performance, at the same time, maintaining speech reconstruction quality.

%The magnitude spectrogram of a speech signal exhibits distinct patterns in the temporal dimension, and the over-suppression error can be easily interpreted compared to that with a raw speech waveform, we focus on alleviating the over-suppression errors using the magnitude spectrogram. 

In speech synthesis~\cite{yamamoto2019probability,yamamoto2020parallel}, the multi-resolution STFT loss function has been used to construct a speech signal. %, which constitutes a spectrum convergence and a magnitude loss. 
Such loss function has also been adopted in speech enhancement~\cite{defossez2020real} to remove noise signals. In addition, delta features have been widely used in speech processing algorithms, e.g., The Mel Frequency Cepstral Coefficients. In a frequency-domain speech separation study named cuPIT-Grid LSTM~\cite{xu2018single}, the delta features have been introduced in the loss function to ensure the continuity of the speech.

Motivated by the above studies, we introduce a hybrid continuity loss function to train a time-domain system, that consists of a time-domain loss and a frequency-domain loss. We use the widely adopted scale-invariant signal-to-noise ratio (SI-SDR)~\cite{le2019sdr} as the time-domain loss, which works well for time-domain systems in reconstructing raw a speech waveform. We also propose a multi-resolution delta spectrum loss in the frequency-domain, to ensure the temporal continuity of a spectrogram. The over-suppressed time-frequency bins or frames will be penalized based on the value of the preceding and following frames, thus alleviating the over-suppression. We examine the loss function on a state-of-the-art time-domain audio-visual speaker extraction algorithm~\cite{pan2021usev}, experimental results show that the hybrid continuity loss function outperforms the use of a single SI-SDR loss function, in improving the downstream ASR performance, while achieving comparable or better signal quality for clean and noisy 2-speaker mixtures simulated from the YouTube LRS2-BBC dataset~\cite{chung2017lip,afouras2018deep}.

The rest of the paper is organized as follows. In Section~\ref{sec:methodology}, we introduce the speaker extraction framework and the proposed hybrid continuity loss. In Section~\ref{sec:experiment}, we present the experimental setup. In Section~\ref{sec:results}, we discuss the experimental results. Finally, we conclude the study in Section~\ref{sec:conclusion}.

\section{Proposed method}
\label{sec:methodology}

Let $x(\tau)$ be a mixture speech in the time-domain, consisting of the target speech $s(\tau)$, interference speech $b_{i}(\tau)$, and noise $n(\tau)$:
\begin{equation}
    \label{eqa:speaker_extraction}
    x(\tau) = s(\tau) + \sum_{i=1}^{I}b_{i}(\tau) + n(\tau)
\end{equation}
where $i \in \{1,...,I\}$ denotes the index of interference speakers. We would like to estimate $\hat{s}(\tau)$ to approximate $s(\tau)$.

\subsection{Neural architecture}

We use an audio-visual time-domain speaker extraction network, that has a similar architecture to the USEV network~\cite{pan2021usev}. It has a speech encoder, a visual encoder, a mask estimator, and a speech decoder, which is illustrated in Fig.~\ref{fig:network}.

The speech encoder transforms the time-domain mixture speech $x(\tau)$ into a frame-based embedding $X(t)$ in the latent space~\footnote{In this paper, a variable with subscript $\tau$ represents it is a time-domain waveform, while a variable with subscript $t$ represents it is a frame-based embedding.} through a convolutional layer $Conv1D(\cdot)$, which behaves like a frequency analyzer~\cite{luo2019conv}: 
\begin{equation}
    \label{eqa:encoder}
    X(t) = ReLU(Conv1D(x(\tau), C, L, L/2))
\end{equation}
where $C=256$, $L=40$, and $L/2$ are the channel, kernel, and stride sizes respectively, $ReLU(\cdot)$ is rectified linear activation.

The visual encoder encodes the target's lip image sequence $v(t)$ into viseme embeddings $V(t)$, which are temporally synchronized with the target's speech in $X(t)$. The $V(t)$ serves as the speaker cue, i.e., lip motion, to attract the target speaker in the following processes. It consists of a ResNet18~\cite{afouras2018conversation}, and 5 layers of temporal convolutional neural network for adaptation~\cite{pan2020multi,wu2019time,reentry}. The ResNet18 is pre-trained on the visual speech recognition task~\cite{afouras2018deep}, and its weights are fixed during the speaker extraction training, to retain the pre-trained knowledge in extracting the lip motion.

The mask estimator takes in the concatenated viseme embeddings $V(t)$ and mixture speech embeddings $X(t)$ as input, to estimate a mask $M(t)$ for the target speech in $X(t)$. The mask estimator has $R=6$ repeated dual-path long short-term memory structures to capture the global dependencies of a speech signal~\cite{luo2020dual}. The mask is element-wise multiplied with $X(t)$ to obtain the extracted speech embeddings $\hat{S}(t)$ in the latent space:
\begin{equation}
    \hat{S}(t) = M(t) \otimes X(t)
\end{equation}

The decoder renders the extracted speech embeddings $\hat{S}(t)$ back to the time-domain speech waveform $\hat{s}(\tau)$. It consists of a linear layer $Linear(\cdot)$ with an overlap-and-add operation $OnA(\cdot)$~\cite{oppenheim1978theory}.
\begin{equation}
    \label{eqa: decoder}
    \hat{s}(\tau) = OnA(Linear(\hat{S}(t), C, L), L/2)
\end{equation}

\subsection{Hybrid loss}

In time-domain end-to-end speaker extraction network training, the negative SI-SDR loss function~\cite{le2019sdr} has been widely used, which is formulated as follow:
\begin{equation}
    \label{eqa:loss_sisnr}
    \mathcal{L}_{SI\mbox{-}SDR} = - 20 \log_{10} \frac{||\frac{<\hat{s},s>s}{||s||^2}||}{||\hat{s} - \frac{<\hat{s},s>s}{||s||^2}||}
\end{equation}
we omit the subscript $\tau$ in the equations for brevity.

We propose a hybrid continuity loss function that consists of the time-domain SI-SDR loss, with a frequency-domain multi-resolution delta spectrum loss as follows:
\begin{equation}
\mathcal{L}_{hybrid}=\mathcal{L}_{SI\mbox{-}SDR}+\gamma \frac{1}{M}\sum^M_{m=1}\mathcal{L}_{freq\mbox{-}\Delta}^m
\end{equation}
where $M=3$ is the number of delta spectrum loss $\mathcal{L}_{freq\mbox{-}\Delta}$ at different resolutions, with number of FFT bins $\in$ \{512, 1024, 2048\},  hop sizes $\in$ \{50, 120, 240\}, and window lengths $\in$ \{240, 600, 1200\} respectively. $\gamma$ is a balancing weight that is set to 1 in this paper.

Each delta spectrum loss consists of a spectrum convergence loss and magnitude loss with delta features, i.e., differential and acceleration features, that is formulated as follow:
\begin{equation}
\mathcal{L}_{freq\mbox{-}\Delta}=\mathcal{L}_{sc\mbox{-}\Delta}+\mathcal{L}_{mag\mbox{-}\Delta}
\end{equation}
\begin{equation}
    \begin{split}
        \mathcal{L}_{sc\mbox{-}\Delta}
        &= \frac{|| \; |STFT(s)| - |STFT(\hat{s})| \; ||_F}{|| \; |STFT(s)| \; ||_F} \\
        &+ \frac{|| \; f_{D}(|STFT(s)|) - f_{D}(|STFT(\hat{s})|) \; ||_F}{|| \; f_{D}(|STFT(s)|) \; ||_F} \\
        &+ \frac{|| \; f_{A}(|STFT(s)|) - f_{A}(|STFT(\hat{s})|) \; ||_F}{|| \; f_{A}(|STFT(s)|) \; ||_F} \\
    \end{split}
\end{equation}
\begin{equation}
    \begin{split}
        \mathcal{L}_{mag\mbox{-}\Delta}
        &= \frac{1}{N} || \; log|STFT(s)| - log|STFT(\hat{s})| \; ||_1 \\
        &+ \frac{1}{N} || \; f_{D}(log|STFT(s)|) - f_{D}(log|STFT(\hat{s})|) \; ||_1 \\ 
        &+ \frac{1}{N} || \; f_{A}(log|STFT(s)|) - f_{A}(log|STFT(\hat{s})|) \; ||_1 \\ 
    \end{split}
\end{equation}
where $STFT(\cdot)$ represents the operation to obtain the magnitude spectrogram of a signal, $N$ denotes the total number of time-frequency bins, $||\cdot||_F$ and $||\cdot||_1$ are the Frobenius and the $L_1$ norms respectively. The function $f_D(\cdot)$ to compute the differential feature is as follow:
\begin{equation}
f_{D}(v(t))=\frac{\sum^L_{l=1} l \times (v(t+l) - v(t-l))}{\sum^L_{l=1} 2l^2}
\end{equation}
where L is the order and sets to $2$ in this paper. The function $f_A(\cdot)$ to compute the acceleration feature is the computation of $f_D(\cdot)$ twice.

With multiple resolutions and the delta features for the frequency-domain loss, we aim to enhance the continuity of the extracted speech magnitude spectrogram with a greater temporal receptive field, and ensure that the over-suppressed frames are seen and compensated by the preceding or following frames, thus reducing over-suppression. In the meantime, the extracted speech is also constrained by the SI-SDR loss, which directly optimizes the extracted signal quality. We expect the hybrid continuity loss function to perform well for both signal quality evaluations and ASR evaluations.

\section{Experimental setup}
\label{sec:experiment}

\subsection{Dataset}
We simulate a 2-speaker mixture dataset named LRS2-mix dataset, from the Oxford-BBC Lip Reading Sentences 2 (LRS2) dataset~\cite{chung2017lip,afouras2018deep} to evaluate the speaker extraction and downstream ASR performances. The LRS2 dataset is an audio-visual dataset, that consists of 45,839, 1,082, and 1,243 clean speech utterances in the original train, validation, and test sets respectively. The speech signal is available at 16 kHz. The face track video of each speech utterance is provided and is available at 25 frames per second. 

To simulate the LRS2-mix dataset, we use the original train, validation, and test sets of the LRS2 dataset to simulate 200,000, 5,000, and 3,000 mixture speech utterances in our train, validation, and test sets respectively. The interference speech is truncated or zero-padded to the length of the target speech first and mixed with the target speech at a random Signal-to-Noise ratio (SNR) set between $10$dB and $-10$dB.

We also simulate a noisy version of the LRS2-mix dataset using babble noise from the WHAM! noise dataset~\cite{Wichern2019WHAM}. For each mixture speech utterance of the LRS2-mix dataset in the train, validation, and test set, a noise signal is randomly sampled from the WHAM! train, validation, and test set respectively and added at a random SNR set between $15$ dB to $-5$ dB. The utterances in the test set do not overlap with those in the train and validation sets, which allows us to perform speaker-independent evaluations.

\subsection{Training}
We implement the model with PyTorch\footnote{The data generation and the training scripts are available at \url{https://github.com/zexupan/avse_hybrid_loss.}}. We use the Adam optimizer with an initial learning rate of $0.001$. The learning rate is halved if the best validation loss (BVL) does not improve for 6 consecutive epochs, and the training stops when the BVL does not improve for 10 consecutive epochs. During training, the utterances are truncated to $6$ seconds to fit into the GPU memory, while the full utterance is used for evaluation.

\subsection{Evaluation metric}

\subsubsection{Signal quality and intelligibility}
We use the SI-SDR and signal-to-noise ratio (SDR) to measure the extracted speech quality as compared with the mixture speech. We use the Perceptual Evaluation of Speech Quality (PESQ) and Short Term Objective Intelligibility (STOI) to measure the perceptual quality and intelligibility of the extracted speech as compared with the mixture speech. The higher the better for the four metrics.

\subsubsection{Automatic speech recognition}
We connect the audio-only or audio-visual \textit{Transformer Connectionist Temporal Classification} TM-CTC model~\cite{afouras2018deep} at the end of the speaker extraction model, to transcribe the extracted speech utterances. The TM-CTC models are pre-trained on the original LRS2 dataset~\cite{chung2017lip,afouras2018deep} that consists of clean speech only~\footnote{The ASR pre-trained models are taken from \url{https://github.com/lordmartian/deep_avsr}}. We use the Word Error Rate (WER) and Character Error Rate (CER) to evaluate the downstream ASR performance. The lower the better for the two metrics.

\subsubsection{Over-suppression}
We use the mean absolute error (MAE) to measure the difference of the magnitude spectrogram between the extracted speech utterance $\hat{s}(\tau)$ and the clean speech utterance $s(\tau)$. The over-suppression MAE and under-suppression MAE for an extracted speech utterance are:
\begin{equation}
MAE_{over}=\frac{1}{N} || \; ReLU( |STFT(s)| - |STFT(\hat{s})| ) \; ||_1
\end{equation}
\begin{equation}
MAE_{under}=\frac{1}{N} || \; ReLU( |STFT(\hat{s})| - |STFT(s)|) \;  ||_1
\end{equation}
where the STFT operation here has FFT bins, hop size, and window length of 512, 120, and 600 respectively. The $ReLU$ operation sets the time-frequency bins to zero if the latter has a negative value.

\begin{table*}
    \centering
    \caption{We report the performance of the time-domain audio-visual speaker extraction network using SI-SDR loss function and our proposed hybrid continuity loss function on the clean and noisy LRS2-mix dataset.}
    \vspace*{-2mm}
    \addtolength{\tabcolsep}{-1pt}
    \begin{tabular}{c|c|c|c|c|c|c|c|c|c|c|c}
        \toprule
        \multirow{1}*{Noise}   &\multirow{2}*{Model}   &\multirow{2}*{SI-SDR}   &\multirow{2}*{SDR}  &\multirow{2}*{PESQ}   &\multirow{2}*{STOI}   &\multicolumn{2}{c|}{Audio-visual ASR}    &\multicolumn{2}{c|}{Audio-only ASR}
        &\multicolumn{2}{c}{MAE}\\
        condition&&   & &&    &\;WER\;  &CER   &\;WER\;  &CER &over   &under\\
        \midrule
        \multirow{4}*{Clean}  
        &Mixture speech           &0.79	&1.10	&1.985	&0.708	&59.2	&36.5	&80.9	&54.7   &-  &-  \\
        &SI-SDR loss~\cite{pan2020muse}        &13.41	&14.08	&2.997	&0.931	&17.3	&7.4	&23.5	&11.3   &0.105	&0.055  \\
        &Hybrid loss (Ours)       &13.48	&14.12	&3.163	&0.934	&16.1	&6.8	&21.2	&10.0   &0.096	&0.060\\
        &Clean speech             &88.12	&288.88	&4.500	&1.000	&11.4	&4.4	&12.5	&4.9    &-  &-  \\
        \midrule
        \multirow{4}*{Noisy}
        &Mixture speech           &-4.43	&-3.81	&1.467	&0.559	&56.5	&32.7	&89.5	&60.6   &-  &-  \\
        &SI-SDR loss~\cite{pan2020muse}        &4.54	&5.91	&2.179	&0.776	&38.8	&20.7	&57.9	&34.6   &0.241	&0.118\\
        &Hybrid loss  (Ours)      &5.12	&6.52	&2.276	&0.789	&34.5	&17.7	&50.8	&29.5   &0.223	&0.119\\
        &Clean speech             &88.12	&288.88	&4.500	&1.000	&11.4	&4.4	&12.5	&4.9    &-  &-  \\
        \bottomrule
    \end{tabular}
    \vspace*{-2mm}
    \addtolength{\tabcolsep}{1pt}
    \label{table:performance}
\end{table*}

\section{Results}
\label{sec:results}

\begin{table}
    \centering
    \caption{Study on the multi-resolution delta spectrum loss. $\Delta$ here means the use of the delta features, i.e., differential and acceleration features. The systems (Sys) here are trained and evaluated on the clean LRS2-mix dataset.}
    \vspace*{-2mm}
    \addtolength{\tabcolsep}{-2pt}
    \begin{tabular}{c|c|c|c|c|c} 
       \toprule
        Sys  &STFT loss   &FFT size     &$\Delta$       &SI-SDR   &WER \\  
        \midrule
        1   &$\mathcal{L}_{sc}+\mathcal{L}_{mag}$
            &\multirow{3}*{\{512,1024,2048\}}
            &\multirow{3}*{\cmark}
                                                     &13.48     &\textbf{16.1}   \\
        2   &$\mathcal{L}_{sc}$                     &&&13.51    &17.0   \\
        3   &$\mathcal{L}_{mag}$                    &&&13.28    &16.5   \\     
        \midrule
        4   &$\mathcal{L}_{sc}+\mathcal{L}_{mag}$
            &\multirow{3}*{\{512,1024,2048\}}
            &\multirow{3}*{\xmark}
                                                      &13.46    &16.3   \\
        5   &$\mathcal{L}_{sc}$                     &&&13.42    &17.1   \\
        6   &$\mathcal{L}_{mag}$                    &&&13.44    &16.6   \\   
        
       \midrule
        7   &\multirow{3}*{$\mathcal{L}_{sc}+\mathcal{L}_{mag}$}  
                     &512                 &\multirow{3}*{\cmark} &13.25          &16.5   \\
        8            &&1024               &&13.42          &17.1   \\
        9            &&2048               &&13.26          &16.3   \\
        
       \bottomrule
    \end{tabular}
    \vspace*{-2mm}
    \addtolength{\tabcolsep}{2pt}
    \label{table:breakdown}
\end{table}

\subsection{Comparison with baseline}
In Table~\ref{table:performance}, we compare the hybrid continuity loss function with the SI-SDR loss function for the time-domain audio-visual speaker extraction network that is shown in Fig.~\ref{fig:network}. We also present the results when the mixture speech and clean speech are used to calculate the metrics.

When the systems are trained and evaluated on the clean LRS2-mix dataset, the hybrid loss function performs similarly to SI-SDR loss for SI-SDR, SDR, and STOI evaluations, but outperforms the latter on PESQ evaluation. The hybrid loss has lower WER and CER for both audio-visual and audio-only ASR evaluations, credited to the 8.6\% reduction of over-suppression MAE. When the systems are trained and evaluated on the noisy LRS2-mix dataset, the hybrid loss outperforms the SI-SDR loss for both the signal quality evaluations, i.e., SI-SDR, SDR, PESQ, STOI, and ASR evaluations, i.e., WER and CER. For both clean and noisy evaluations, the MAE for over-suppression decreases, but that for under-suppression increases, which further shows that the ASR algorithm tolerates more under- than over-suppression errors.

\subsection{Ablation studies}
Next, we study the contribution of individual components from the multi-resolution delta spectrum loss and present the results in Table~\ref{table:breakdown}. System 1 is trained with our proposed hybrid continuity loss function, while systems 2-9 are trained with the SI-SDR loss together with part of the multi-resolution delta spectrum loss. We report the SI-SDR and the audio-visual WER here. All systems achieve similar SI-SDR, which is similar to using the SI-SDR loss function alone, i.e., 13.41 in Table~\ref{table:performance}, this shows that the use of the frequency-domain loss does not harm the signal reconstruction from 2-speaker mixture speech.

System 2 only has the spectrum convergence loss in the hybrid continuity loss function, while system 3 only has that of the magnitude loss. Both systems are outperformed by system 1 on WER evaluations, which used our proposed loss. Both the spectrum convergence loss and the magnitude loss contribute to the reduction of WER. System 4 does not include the delta features, i.e., differential and acceleration features, the WER increases by 0.2 \% compared with system 1, which shows that the delta features are useful. Systems 7-9 only use single-resolution for STFT in the hybrid continuity loss function, none of them outperform system 1 in terms of WER, which uses that of multi-resolution loss, which shows the importance of various resolutions.

We also present the SI-SDR and the audio-visual WER evaluations of the baseline and the proposed system on the clean LRS2-mix dataset with various target-interference SNR in Figures 2 and 3. The target-interference SNR is defined as the energy contrast between the target speaker and the interference speaker in the mixture speech in terms of SNR. As the input mixture becomes less noisy, i.e. higher SNR, the SI-SDR increases and the WER decreases. We observe that the improvement in WER is less affected by the target-interference SNR than the SI-SDR. The hybrid loss outperforms the SI-SDR loss in terms of WER evaluation and achieves a similar result on the SI-SDR evaluation.

\begin{figure}[t]
\begin{minipage}[t]{.49\linewidth}
  \centering
  \centerline{\includegraphics[width=\linewidth]{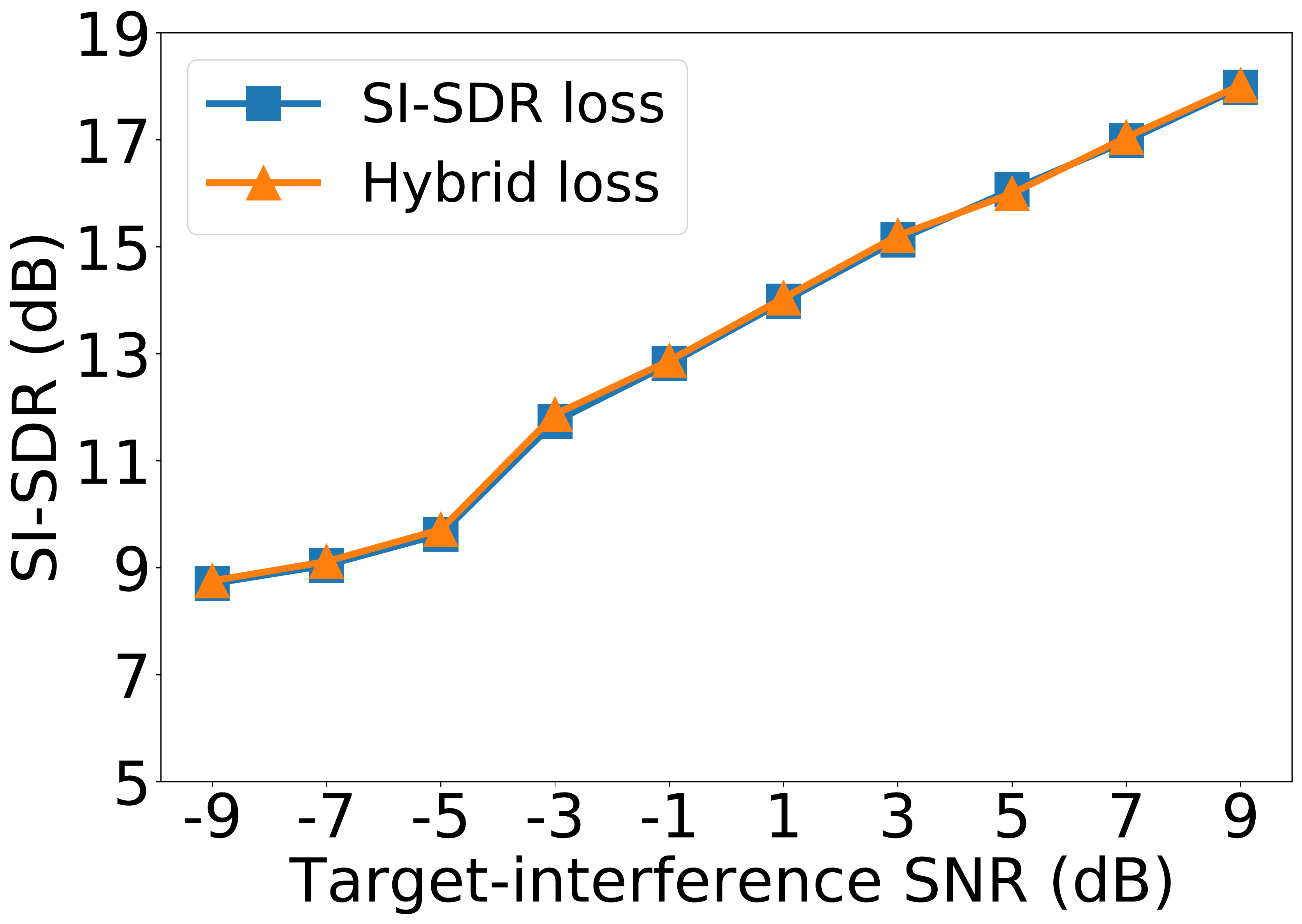}}
  \vspace*{-2mm}
  \caption{The SI-SDR against the target-interference SNR.}\medskip
  \label{fig:sisdr}
\end{minipage}
\hfill
\begin{minipage}[t]{.49\linewidth}
  \centering
  \centerline{\includegraphics[width=\linewidth]{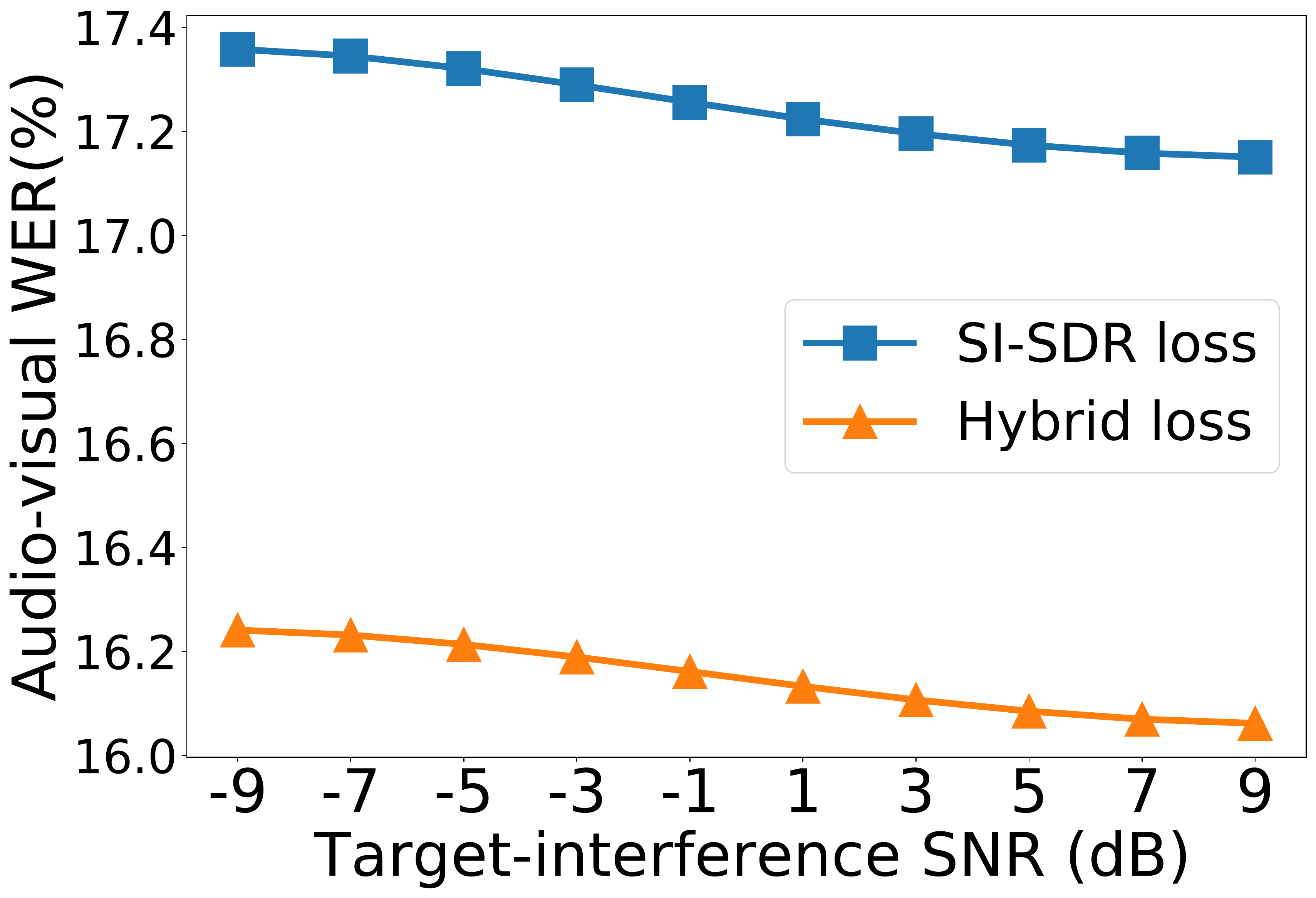}}
  \vspace*{-2mm}
  \caption{The WER against the target-interference SNR.}\medskip
  \label{fig:wer}
\end{minipage}
\vspace*{-5mm}
\end{figure}

\section{Conclusion}
\label{sec:conclusion}
In this work, we proposed a hybrid continuity loss function that consists of an SI-SDR loss and a multi-resolution delta spectrum loss, to alleviate the over-suppression errors in time-domain speaker extraction networks. This makes a step forward for downstream ASR algorithms.

% \section{Acknowledgement}
% This research is supported by The Science and Engineering Research Council, Agency for Science, Technology and Research (A*STAR), Singapore, through the National Robotics Program under Human-Robot Interaction Phase 1 (Grant No. 192 25 00054).  This research is also funded by the Deutsche Forschungsgemeinschaft (DFG, German Research Foundation) under Germany's Excellence Strategy (University Allowance, EXC 2077, University of Bremen), and by The Chinese University of Hong Kong, Shenzhen (UDF01002333, UF02002333).

\bibliographystyle{IEEEtran}
\bibliography{mybib}

\begin{thebibliography}{10}
\providecommand{\url}[1]{#1}
\csname url@samestyle\endcsname
\providecommand{\newblock}{\relax}
\providecommand{\bibinfo}[2]{#2}
\providecommand{\BIBentrySTDinterwordspacing}{\spaceskip=0pt\relax}
\providecommand{\BIBentryALTinterwordstretchfactor}{4}
\providecommand{\BIBentryALTinterwordspacing}{\spaceskip=\fontdimen2\font plus
\BIBentryALTinterwordstretchfactor\fontdimen3\font minus
  \fontdimen4\font\relax}
\providecommand{\BIBforeignlanguage}[2]{{%
\expandafter\ifx\csname l@#1\endcsname\relax
\typeout{** WARNING: IEEEtran.bst: No hyphenation pattern has been}%
\typeout{** loaded for the language `#1'. Using the pattern for}%
\typeout{** the default language instead.}%
\else
\language=\csname l@#1\endcsname
\fi
#2}}
\providecommand{\BIBdecl}{\relax}
\BIBdecl

\bibitem{bronkhorst2000cocktail}
A.~W. Bronkhorst, ``The cocktail party phenomenon: A review of research on
  speech intelligibility in multiple-talker conditions,'' \emph{Acta Acustica
  united with Acustica}, vol.~86, no.~1, pp. 117--128, 2000.

\bibitem{ephrat2018looking}
A.~Ephrat, I.~Mosseri, O.~Lang, T.~Dekel, K.~Wilson, A.~Hassidim, W.~T.
  Freeman, and M.~Rubinstein, ``Looking to listen at the cocktail party: a
  speaker-independent audio-visual model for speech separation,'' \emph{ACM
  Trans. Graph.}, vol.~37, no.~4, pp. 1--11, 2018.

\bibitem{reentry}
Z.~Pan, R.~Tao, C.~Xu, and H.~Li, ``Selective listening by synchronizing speech
  with lips,'' \emph{IEEE/ACM Trans. Audio, Speech, Lang. Process.}, vol.~30,
  pp. 1650--1664, 2022.

\bibitem{wang2017deep}
D.~Wang, ``Deep learning reinvents the hearing aid,'' \emph{IEEE spectrum},
  vol.~54, no.~3, pp. 32--37, 2017.

\bibitem{tao2021someone}
R.~Tao, Z.~Pan, R.~K. Das, X.~Qian, M.~Z. Shou, and H.~Li, ``Is someone
  speaking? {E}xploring long-term temporal features for audio-visual active
  speaker detection,'' in \emph{Proc. of the 29th ACM Int. Conf. on
  Multimedia}, 2021, pp. 3927--3935.

\bibitem{yue2019end}
X.~Yue, G.~Lee, E.~Y{\i}lmaz, F.~Deng, and H.~Li, ``End-to-end code-switching
  {ASR} for low-resourced language pairs,'' in \emph{Proc. IEEE Autom. Speech
  Recognit. Understanding Workshop}, 2019, pp. 972--979.

\bibitem{song2022multimodal}
Q.~Song, B.~Sun, and S.~Li, ``Multimodal sparse transformer network for
  audio-visual speech recognition,'' \emph{IEEE Transactions on Neural Networks
  and Learning Systems}, 2022.

\bibitem{qian2021multi}
X.~Qian, M.~Madhavi, Z.~Pan, J.~Wang, and H.~Li, ``Multi-target {DoA}
  estimation with an audio-visual fusion mechanism,'' in \emph{Proc. IEEE Int.
  Conf. Acoust., Speech, Signal Process.}, 2021, pp. 4280--4284.

\bibitem{luo2020dual}
Y.~{Luo}, Z.~{Chen}, and T.~{Yoshioka}, ``{Dual-path RNN: Efficient long
  sequence modeling for time-domain single-channel speech separation},'' in
  \emph{Proc. IEEE Int. Conf. Acoust., Speech, Signal Process.}, 2020, pp.
  46--50.

\bibitem{luo2019conv}
Y.~{Luo} and N.~{Mesgarani}, ``{Conv-TasNet: Surpassing ideal time–frequency
  magnitude masking for speech separation},'' \emph{IEEE/ACM Trans. Audio,
  Speech, Lang. Process.}, vol.~27, no.~8, pp. 1256--1266, 2019.

\bibitem{kolbaek2017multitalker}
M.~Kolb{\ae}k, D.~Yu, Z.-H. Tan, and J.~Jensen, ``Multitalker speech separation
  with utterance-level permutation invariant training of deep recurrent neural
  networks,'' \emph{IEEE/ACM Trans. Audio, Speech, Lang. Process.}, vol.~25,
  no.~10, pp. 1901--1913, 2017.

\bibitem{hershey2016deep}
J.~R. {Hershey}, Z.~{Chen}, J.~{Le Roux}, and S.~{Watanabe}, ``Deep clustering:
  Discriminative embeddings for segmentation and separation,'' in \emph{Proc.
  IEEE Int. Conf. Acoust., Speech, Signal Process.}, 2016, pp. 31--35.

\bibitem{zeghidour2020wavesplit}
N.~Zeghidour and D.~Grangier, ``Wavesplit: End-to-end speech separation by
  speaker clustering,'' \emph{preprint arXiv:2002.08933}, 2020.

\bibitem{liu2019divide}
Y.~Liu and D.~Wang, ``Divide and conquer: A deep {CASA} approach to
  talker-independent monaural speaker separation,'' \emph{IEEE/ACM Trans.
  Audio, Speech, Lang. Process.}, vol.~27, no.~12, pp. 2092--2102, 2019.

\bibitem{wang2020voicefilter}
Q.~Wang, I.~L. Moreno, M.~Saglam, K.~Wilson, A.~Chiao, R.~Liu, Y.~He, W.~Li,
  J.~Pelecanos, M.~Nika \emph{et~al.}, ``{VoiceFilter-Lite}: Streaming targeted
  voice separation for on-device speech recognition,'' \emph{Proc.
  INTERSPEECH}, pp. 2677--2681, 2020.

\bibitem{Chenglin2020spex}
C.~{Xu}, W.~{Rao}, E.~S. {Chng}, and H.~{Li}, ``{SpEx}: Multi-scale time domain
  speaker extraction network,'' \emph{IEEE/ACM Trans. Audio, Speech, Lang.
  Process.}, vol.~28, pp. 1370--1384, 2020.

\bibitem{spex_plus2020}
M.~Ge, C.~Xu, L.~Wang, E.~S. Chng, J.~Dang, and H.~Li, ``{SpEx}+: A complete
  time domain speaker extraction network,'' in \emph{Proc. INTERSPEECH}, 2020,
  pp. 1406--1410.

\bibitem{xiao2019single}
X.~Xiao, Z.~Chen, T.~Yoshioka, H.~Erdogan, C.~Liu, D.~Dimitriadis, J.~Droppo,
  and Y.~Gong, ``Single-channel speech extraction using speaker inventory and
  attention network,'' in \emph{Proc. IEEE Int. Conf. Acoust., Speech, Signal
  Process.}, 2019, pp. 86--90.

\bibitem{shi2020speaker}
J.~Shi, J.~Xu, Y.~Fujita, S.~Watanabe, and B.~Xu, ``Speaker-conditional chain
  model for speech separation and extraction,'' in \emph{Proc. INTERSPEECH},
  2020, pp. 2707--2711.

\bibitem{delcroix2020improving}
M.~Delcroix, T.~Ochiai, K.~Zmolikova, K.~Kinoshita, N.~Tawara, T.~Nakatani, and
  S.~Araki, ``Improving speaker discrimination of target speech extraction with
  time-domain {SpeakerBeam},'' in \emph{Proc. IEEE Int. Conf. Acoust., Speech,
  Signal Process.}, 2020, pp. 691--695.

\bibitem{wu2019time}
J.~{Wu}, Y.~{Xu}, S.~{Zhang}, L.~{Chen}, M.~{Yu}, L.~{Xie}, and D.~{Yu}, ``Time
  domain audio visual speech separation,'' in \emph{Proc. IEEE Autom. Speech
  Recognit. Understanding Workshop}, 2019, pp. 667--673.

\bibitem{afouras2018conversation}
T.~Afouras, J.~S. Chung, and A.~Zisserman, ``The conversation: Deep
  audio-visual speech enhancement,'' in \emph{Proc. INTERSPEECH}, 2018, pp.
  3244--3248.

\bibitem{ochiai2019multimodal}
T.~Ochiai, M.~Delcroix, K.~Kinoshita, A.~Ogawa, and T.~Nakatani, ``{Multimodal
  SpeakerBeam}: Single channel target speech extraction with audio-visual
  speaker clues,'' in \emph{Proc. INTERSPEECH}, 2019, pp. 2718--2722.

\bibitem{seg_pan}
Z.~Pan, X.~Qian, and H.~Li, ``Speaker extraction with co-speech gestures cue,''
  \emph{IEEE Signal Processing Letters}, 2022.

\bibitem{koichiro2021}
K.~Ito, M.~Yamamoto, and K.~Nagamatsu, ``Audio-visual speech enhancement method
  conditioned in the lip motion and speaker-discriminative embeddings,'' in
  \emph{Proc. IEEE Int. Conf. Acoust., Speech, Signal Process.}, 2021, pp.
  6668--6672.

\bibitem{michelsanti2021overview}
D.~Michelsanti, Z.-H. Tan, S.-X. Zhang, Y.~Xu, M.~Yu, D.~Yu, and J.~Jensen,
  ``An overview of deep-learning-based audio-visual speech enhancement and
  separation,'' \emph{IEEE/ACM Trans. Audio, Speech, Lang. Process.}, 2021.

\bibitem{pan2020muse}
Z.~Pan, R.~Tao, C.~Xu, and H.~Li, ``{MuSE}: Multi-modal target speaker
  extraction with visual cues,'' in \emph{Proc. IEEE Int. Conf. Acoust.,
  Speech, Signal Process.}, 2021, pp. 6678--6682.

\bibitem{pan2021usev}
Z.~Pan, M.~Ge, and H.~Li, ``{USEV}: Universal speaker extraction with visual
  cue,'' \emph{arXiv preprint arXiv:2109.14831}, 2021.

\bibitem{yamamoto2019probability}
R.~Yamamoto, E.~Song, and J.-M. Kim, ``Probability density distillation with
  generative adversarial networks for high-quality parallel waveform
  generation,'' \emph{Proc. INTERSPEECH}, pp. 699--703, 2019.

\bibitem{yamamoto2020parallel}
------, ``Parallel wavegan: A fast waveform generation model based on
  generative adversarial networks with multi-resolution spectrogram,'' in
  \emph{Proc. IEEE Int. Conf. Acoust., Speech, Signal Process.}, 2020, pp.
  6199--6203.

\bibitem{defossez2020real}
A.~D{\'e}fossez, G.~Synnaeve, and Y.~Adi, ``Real time speech enhancement in the
  waveform domain,'' \emph{Proc. INTERSPEECH}, pp. 3291--3295, 2020.

\bibitem{xu2018single}
C.~Xu, W.~Rao, X.~Xiao, E.~S. Chng, and H.~Li, ``Single channel speech
  separation with constrained utterance level permutation invariant training
  using grid {LSTM},'' in \emph{Proc. IEEE Int. Conf. Acoust., Speech, Signal
  Process.}, 2018, pp. 6--10.

\bibitem{le2019sdr}
J.~Le~Roux, S.~Wisdom, H.~Erdogan, and J.~R. Hershey, ``{SDR}--half-baked or
  well done?'' in \emph{Proc. IEEE Int. Conf. Acoust., Speech, Signal
  Process.}, 2019, pp. 626--630.

\bibitem{chung2017lip}
J.~S. Chung, A.~Senior, O.~Vinyals, and A.~Zisserman, ``Lip reading sentences
  in the wild,'' in \emph{IEEE conference on computer vision and pattern
  recognition}, 2017, pp. 3444--3453.

\bibitem{afouras2018deep}
T.~Afouras, J.~S. Chung, A.~Senior, O.~Vinyals, and A.~Zisserman, ``Deep
  audio-visual speech recognition,'' \emph{IEEE Trans. Pattern Anal. Mach.
  Intell}, 2018.

\bibitem{pan2020multi}
Z.~Pan, Z.~Luo, J.~Yang, and H.~Li, ``Multi-modal attention for speech emotion
  recognition,'' in \emph{Proc. INTERSPEECH}, 2020, pp. 364--368.

\bibitem{oppenheim1978theory}
A.~Oppenheim and R.~Schafer, ``Theory and application of digital signal
  processing,'' \emph{Englewood Cliffs}, 1978.

\bibitem{Wichern2019WHAM}
G.~Wichern, J.~Antognini, M.~Flynn, L.~R. Zhu, E.~McQuinn, D.~Crow, E.~Manilow,
  and J.~Le~Roux, ``Wham!: Extending speech separation to noisy environments,''
  in \emph{Proc. INTERSPEECH}, 2019.

\end{thebibliography}

\end{document}